\RequirePackage{fix-cm}
\documentclass[smallextended]{svjour3}       
\smartqed  
\usepackage{graphicx}
\usepackage{amssymb}
\usepackage{mathptmx}      
\usepackage{latexsym}
\begin{document}

\title{Some clarifications about the Bohmian geodesic deviation equation and Raychaudhuri's equation}

\author{Faramarz Rahmani\and Mehdi Golshani}
\institute{F. Rahmani \at
              School of Physics, Institute for Research in Fundamental Science(IPM), Tehran, Iran\\
              Tel.: +98-21-22180692,
              Fax: +98-21-22280415\\
              \email{faramarzrahmani@ipm.ir}           
           \and
           M. Golshani \at
              School of Physics, Institute for Research in Fundamental Science(IPM), Tehran, Iran\\
              Department of Physics, Sharif University of Technology, Tehran, Iran\\
              Tel.:+98-21-66022718, Fax.:+98-21-66022718\\
              \email{golshani@sharif.edu}
}              
\date{Received: date / Accepted: date}

\maketitle

\begin{abstract}
One of the important and famous topics in general theory of relativity and gravitation is the problem of geodesic deviation and its related singularity theorems. An interesting subject is the investigation of these concepts when quantum effects are considered. Since, the definition of trajectory is not possible in the framework of standard quantum mechanics (SQM), we investigate the problem of geodesic equation and its related topics in the framework of Bohmian quantum mechanics in which the definition of trajectory is possible. We do this in a fixed background and we do not consider the back-reaction effects of matter on the spacetime metric.
\end{abstract}

\section{Introduction}
\label{sec:1}
Bohmian quantum mechanics (BQM) allows us to attribute trajectories to particles like in classical mechanics. But here there is a difference due to the fact that these trajectories are affected by a wave function or a guiding wave. This encourages us to study Bohmian geodesic deviation equation and Raychaudhuri's equation, where quantum effects are significant. In this paper, we do this for relativistic Klein-Gordon's particles and obtain their non-relativistic limit systematically. But, we first review the elements of BQM rapidly. In ref \cite{RefD} a similar work has been done. Our work has some differences in method and results with respect to the methods and results of ref \cite{RefD}.\par 
First, we explain the postulates of BQM and its differences with SQM in summary. In BQM, the wave function or pilot-wave is defined in the configuration space of the system, but it has a real origin. Its main task is guiding particles on some real trajectories in a causal manner. In other words, the wave function is not merely a probabilistic instrument(as in Born's interpretation of the wave function in SQM). But, since the initial conditions are not provided exactly, we attribute a statistical interpretation to it, with the distribution $\rho=|\psi|^2$. Another postulate of BQM is that we can attribute physical properties like energy, momentum and etc to a particle even before experiment. \cite{RefHolland,Refuniverse}. The experimental results of both approach are identical. In fact, in BQM, like in classical mechanics, a particle has deterministic trajectories. These trajectories have different features from the classical trajectories. Because, they are determined by a wave function which is a quantum concept. In Bohm's own view, the effects of the guiding wave on a system  appears in an important concept known as quantum potential. But in some approaches like the DGZ theory, quantum potential has no essential role\cite{RefGold}. We follow Bohm's own approach in following.\par  By substituting the polar form $\psi(\mathbf{x},t)=R(\mathbf{x},t)\exp (\frac{iS(\mathbf{x},t)}{\hbar})$ of the wave function into the Schr\"{o}dinger equation, we get a generalized Hamilton-Jacobi equation and a continuity equation:
\begin{equation}\label{hamilton11}
\frac{\partial S(\mathbf{x},t)}{\partial t}+\frac{(\nabla S)^2}{2m}+V(\mathbf{x})+Q(\mathbf{x})=0
\end{equation}
\begin{equation}\label{hamilton12}
\frac{\partial R^2}{\partial t}+\frac{1}{m}\mathbf{\nabla}\cdot(R^2 \nabla S)=0
\end{equation}
In the above equations, $S(\mathbf{x},t)$ is the action of the particle and $Q$ is the non-relativistic Bohmian quantum potential defined by:
\begin{equation}\label{potential1}
Q=-\frac{\hbar^2 \nabla^2 R(\mathbf{x},t)}{2mR(\mathbf{x},t)}=-\frac{\hbar^2 \nabla^2 \sqrt{\rho(\mathbf{x},t)}}{2m\sqrt{\rho(\mathbf{x},t)}}
\end{equation}
where, $\rho=R^2$. The quantum potential has some special properties like context dependency and non-locality which give non-classical properties to a particle. \par The position of the particle is obtained from the following equation:
\begin{equation}\label{guidance}
\frac{d\mathbf{x}(t)}{dt}=\left(\frac{\nabla S(\mathbf{x},t)}{m}\right)_{\mathbf{X}=\mathbf{x}(t)}
\end{equation}
where $\nabla S(\mathbf{x},t)$ is the momentum of the particle. By knowing the initial position $\mathbf{x}$ and wave function $\psi(\mathbf{x},t)$, the future of the system is obtained.The expression $\mathbf{X}=\mathbf{x}(t)$ means that among all possible trajectories, in an ensemble of particles, we choose one of them. The relation (\ref{guidance}) is called guidance formula. In ref \cite{RefHolland}, it has been demonstrated that the trajectories following equation (\ref{guidance}) do not cross each other. So they can form a congurences of trajectories. The single-valuedness of the wave function leads to:
\begin{equation}
\Delta S = \oint dS = \oint \nabla S \cdot d\mathbf{x}= \oint \mathbf{p}\cdot d\mathbf{x}=nh
\end{equation}
In other words, the difference between phases are constant and for every instant at the place $\mathbf{x}$, $\mathbf{p}=\nabla S$ is single-valued. The trajectories are irrotational, because, $\nabla \times \mathbf{p}=\nabla \times \nabla S=0$. For getting equations (\ref{hamilton11}),(\ref{hamilton12}) and (\ref{potential1}), Bohm's approach is not the only possible approach. It can also obtain the result through a variational method. (See for example refs \cite{RefAtigh,RefAtigh2,RefRah}). \par For a relativistic spinless particle, we substitute the polar form of the wave function into the Klein-Gordon equation to derive the quantum mechanical Hamilton-Jacobi equation. For a relativistic spinless particle in a curved space-time we have:
\begin{equation}\label{rela}
\nabla_{\mu} S \nabla ^{\mu} S =g^{\mu\nu}\nabla _{\nu}S\nabla_{\mu} S =m_0^2(1+\mathfrak{Q})=\mathcal{M}^2
\end{equation}
where
\begin{equation}\label{repot}
\mathfrak{Q}=\frac{\hbar^2}{m_0^2}\frac{\nabla_\mu \nabla^\mu R}{R}=\frac{\hbar^2}{m_0^2}\frac{\nabla_\mu \nabla^\mu \sqrt{\rho}}{\sqrt{\rho}}
\end{equation}
Equation (\ref{rela}) indicates that the rest mass of the particle is not a constant in the rest frame of the particle; rather, it depends on the quantum potential. The 4-momentum of the particle is obtained from:
\begin{equation}\label{pp}
p^\mu =\mathcal{M}u^\mu =m_0(\sqrt{1+Q})u^\mu =-\nabla^\mu S
\end{equation}
which depends on the quantum potential.\par 
Classical deviation equation for a congurence of geodesics is 
\begin{equation}\label{dev1}
\frac{D^2 \eta^\mu}{D\tau^2}= R^\mu _{\rho \lambda \nu}u^\rho  \eta ^\lambda u^\nu
\end{equation}
where $\eta^\mu$ is the component of deviation vector between two neighborhood geodesics and $\tau$ is the affine parameter of geodesics. Since we only consider timelike geodesics, then $\tau$ is the proper time. The classical Raychaudhuri's equation for a congurence of geodesics is 
\begin{equation}\label{ray1}
\frac{d\theta}{d\tau}=\omega_{\mu \nu} \omega^{\mu \nu}-\sigma_{\mu \nu} \sigma^{\mu \nu}-\frac{1}{3}\theta^2 -R_{\mu \nu}u^\mu u^\nu
\end{equation}
The quantity $\theta= \nabla_\mu u^\mu$ is called expansion, which measures the fractional change of volume per unit time. The shear tensor is $\sigma_{\mu\nu}=\frac{1}{2}(h^\lambda_\nu \nabla_\lambda u_\mu + h^\lambda_\mu \nabla_\lambda u_\nu)-\frac{1}{3}\theta h_{\mu \nu}$, where $h_{\mu \nu}$ is the transverse metric and $h^\mu_\nu$ is projection operator constructed from it. The shear tensor is symmetric and trace-free. Also, the rotation tensor is $\omega_{\mu \nu}=\frac{1}{2}(h^\lambda_\nu \nabla_\lambda u_\mu - h^\lambda_\mu \nabla_\lambda u_\nu)$, which is an antisymmetric tensor. Both shear and rotation are orthogonal to the four-velocity $u^\mu$ i.e, $\omega_{\mu\nu}u^\nu = \sigma_{\mu\nu}u^\nu =0$. \par In general theory of relativity, there is an important theorem known as focusing theorem which states that for hypersurface orthogonal geodesics $(\omega_{\mu\nu}=0)$, with a strong energy condition $R_{\mu \nu}u^\mu u^\nu >0$, the geodesics get focused and converge to a caustic point. This occurs in finite proper time $\tau \leq \frac{3}{\theta(0)}$ where, $\theta(0)$ is the initial expansion \cite{RefWald,RefCar,RefP,RefPad}. In the next sections, we investigate these topics in the framework of Bohmian mechanics. \par 
As we mentioned before, in ref \cite{RefD} a similar work has been done, using the velocity field approach. The author, has began from the non-relativistic case and followed its consequences in the relativistic case. But here, we begin from a different starting point and go to the non-relativistic domain systematically. The mathematical result of our approach is somehow different from the result of ref \cite{RefD} in relativistic domain. But it coincides with it in the non-relativistic domain. 
We do it for the minimal conformal coupling of the Klein-Gordon equation, i.e, $\zeta=0$ in equation $(\Box + m^2 -\zeta \mathcal{R})\phi=0$, where $\mathcal{R}$ is Ricci scalar. The extension for non-zero $\zeta$ is straightforward. Here, we fix the background metric and do not consider the back-reaction effects. The back-reaction effects of matter on the space-time metric, in the framework of BQM, has been studied in ref \cite{Reffatima}. In this work, we choose the Lorentzian signature, as ($+1, -1,-1,-1$).

\section{Deriving Bohmian geodesic deviation equation and Raychaudhuri's equation }\label{sec:2}
The main key that helps us to obtain Bohmian versions of deviation equation and Raychaudhari's equation is that when we consider the trajectory of a particle in the presence of a quantum force, the geodesic equation $u^\nu \nabla_\nu u^\mu=0$ can not be valid anymore. In fact, the particle on such terajectory experiences an acceleration as $a^\mu = u^\nu \nabla_\nu u^\mu$. From equations (\ref{rela}) and (\ref{pp}) we obtain:
\begin{equation}\label{imp}
p^\mu p_\mu =\mathcal{M}^2
\end{equation}
This ensures that:
\begin{equation}\label{u}
u^\mu u_\mu = g_{\mu \nu} u^\mu u^\nu =1 
\end{equation}
This is a good news, because the geodesics are timelike and the classical treatment of gravity for constructing geometrical instrument for studying geodesic deviation can be continued. In other words, we can evaluate the essential quantities like expansion, shear, rotation and etc, by using a purely spatial transverse metric $h_{\mu\nu}=g_{\mu\nu}-u_\mu u_\nu$. Also, we have $h_{\mu\nu}u^\nu =0$ and $\nabla_\lambda h_{\mu \nu}=0$. It is noteworthy that the momentum is defined as $p^\mu=\mathcal{M}u^\mu$, and not as $p^\mu =m_0 u^\mu$. If the relation (\ref{u}) is not be true, we can not construct projection operator and the whole of procedure faces with some difficulties. As we mentioned earlier, we should only consider an acceleration in addition to the classical situation. This acceleration is related to Bohm's quantum potential. 
The relation (\ref{imp}) is a general relation which is true for spinning particles. Naturally, for spinning particles $\mathcal{M}^2$ should contain spin four-vector and its derivatives. For the study of the quantum potential for spinning particles, refs \cite{RefHolland} and \cite{RefHC,RefDoran,RefHest} are suitable. In fact $\mathcal{M}^2=m_0^2 e^{\mathfrak{Q}}$ which is derivable from Mach's principle or by taking mass as a dynamical field \cite{RefBrans,Refshoja2}. But since, the quantum mechanical energies are very small relative to the classical energies, we can write $e^{\mathfrak{Q}}\simeq 1+\mathfrak{Q}$.
Now, from the equation (\ref{imp}) we have:
\begin{eqnarray}\label{ac}
2p_\mu \frac{dp^\mu}{d\tau}&=&2\mathcal{M}\frac{d\mathcal{M}}{d\tau} \nonumber \\
(u^\nu \nabla_\nu p^\mu)p_\mu &=&\mathcal{M} u_\mu \nabla^\mu \mathcal{M}\nonumber \\
u^\nu \nabla_\nu (\mathcal{M} u^\mu)&=&\nabla ^\mu \mathcal{M}\nonumber \\
\mathcal{M}u^\nu \nabla_\nu u^\mu &=&-(u^\nu\nabla_\nu \mathcal{M})u^\mu +\nabla^\mu \mathcal{M}\nonumber \\
u^\nu \nabla_\nu u^\mu &=& -\frac{1}{\mathcal{M}}(u^\nu\nabla_\nu \mathcal{M})u^\mu +\frac{1}{\mathcal{M}}\nabla^\mu \mathcal{M}\nonumber \\
u^\nu \nabla_\nu u^\mu &=& -\frac{1}{2}u^\mu u^\nu \nabla_\nu ln(1+\mathfrak{Q}) +\frac{1}{2}\nabla^\mu ln(1+\mathfrak{Q})=a^\mu
\end{eqnarray}
Infact, $a^\mu$ is the quantum acceleration of the particle.
If we multiply both sides of the above equation by $u_\mu$, we get $a^\mu u_\mu=0$, which indicates that four-acceleration is purely spacelike. Since the energies due to quantum potential $\mathfrak{Q}$, is very small with respect to the classical energies, we can assume that $\mathfrak{Q}<< 1$ and $ln(1+\mathfrak{Q}) \simeq \mathfrak{Q}$, and the last relation reduces to:
\begin{equation}\label{acac}
a^\mu=-\frac{1}{2}u^\mu u^\nu \nabla_\nu \mathfrak{Q}+\frac{1}{2}\nabla^\mu \mathfrak{Q}=-\frac{1}{2}u^\mu u^\nu \nabla_\nu \mathfrak{Q}+\frac{1}{2}g^{\mu\nu}\nabla_\nu \mathfrak{Q}=\frac{1}{2}h^{\mu\nu}\nabla_\nu \mathfrak{Q}\equiv a^\mu _{(B)}
\end{equation}
By $a^\mu _{(B)}$, we mean that this acceleration is due to the Bohmian quantum potential.
Now, by using the relativistic Bohmian quantum potential, we have:
\begin{equation}\label{a4}
a^\mu_{(B)} = -\frac{1}{2}u^\mu u^\nu \nabla_\nu \left(\frac{\hbar^2}{m_0^2}\frac{\nabla_\lambda \nabla^\lambda \sqrt{\rho}}{\sqrt{\rho}}\right)+\frac{1}{2}g^{\mu\nu}\nabla_\nu \left( \frac{\hbar^2}{m_0^2}\frac{\nabla_\lambda \nabla^\lambda \sqrt{\rho}}{\sqrt{\rho}}\right)
\end{equation}
We shall demonstrate that above equation, in the non-relativistic domain, is compatible with an acceleration due to the non-relativistic Bohmian potential.
It is obvious from the relation (\ref{potential1}) that in non-relativistic domain only the spatial derivatives appear in the quantum potential, i.e, $\frac{\partial Q}{\partial t}=0$. Also we have, $u^i =\delta^\mu_0$ and  $\tau \rightarrow t$. We infer from the equation  (\ref{a4}) that the acceleration in the non-relativistic limit is:
\begin{equation}
a^i = \frac{1}{2}\partial^i\left(-\frac{\hbar^2}{m^2_0}\frac{\nabla^2 \sqrt{\rho}}{\sqrt{\rho}}\right)=\partial^i\left(-\frac{\hbar^2}{2m^2_0}\frac{\nabla^2 \sqrt{\rho}}{\sqrt{\rho}}\right)=\frac{\partial^i Q}{m_0}=\frac{f^i}{m_0}
\end{equation}
 where $Q$ is the non-relativistic Bohmian quantum potential and $f^i$ is non-relativistic quantum force. For simplicity, we have considered the last relation in a flat spacetime. Otherwise, a gravitational acceleration $\nabla \phi(\mathbf{x})$ should have been added to it.  \par 
Now, we consider the effect of such acceleration on the deviation equation. Then we derive the modified Raychaudhari's equation. Finally, we shall discuss about their consequences.
\subsection{Bohmian deviation equation}
Consider a family of non-crossing  curves which are not geodesics in general. The affine parameter along a specified curve is $\tau$ and tangent to it is the vector $u^\mu=\frac{\partial x^\mu}{\partial \tau}$. The deviation vector between two neighboring curves is defined as $\eta^\mu=\frac{\partial x^\mu }{\partial s} $ in which the parameter $ s$ parameterizes the curves in which $\eta^\mu$ is tangent to them. Also, $\eta^\mu$ and $u^\mu$ are orthogonal. The velocity field for deviation vector is defined as $\frac{d\eta^\mu}{d\tau}= u^\nu \nabla_\nu \eta^\mu$. The acceleration of deviation vector which we are interested in it, is
\begin{equation}
\frac{D^2 \eta^\mu}{D\tau^2}=\nabla_\lambda (u^\nu \nabla_\nu \eta^\mu)u^\lambda = \nabla_\lambda (\eta^\nu \nabla_\nu u^\mu)u^\lambda 
\end{equation}
where we have used from this fact that: $\frac{\partial u^\mu}{\partial s}= \frac{\partial \eta^\mu }{\partial \tau}$. After some calculations and using the definition of curvature tensor for any arbitrary vector field, $\left[\nabla_\rho \nabla_\lambda - \nabla_\lambda \nabla_\rho \right]A^\mu =R^\mu_{\nu\rho\lambda}A^\nu $, we get:
\begin{equation}
\frac{D^2 \eta^\mu}{D\tau^2}=\eta^\lambda \nabla_\lambda(u^\nu \nabla_\nu u^\mu)+R^\mu _{\rho \lambda \nu}u^\rho  \eta ^\lambda u^\nu 
\end{equation}
For a geodesic equation, the first term vanishes. But, we keep the first term because we deal with the non-geodesics curves due to the presence of Bohmian quantum force. So, we have:
\begin{equation}\label{aa3}
\frac{D^2 \eta^\mu}{D\tau^2}=\eta^\lambda \nabla_\lambda a^\mu _{(B)} + R^\mu _{\rho \lambda \nu}u^\rho  \eta ^\lambda u^\nu .
\end{equation}
This equation can be written in an interesting form:
\begin{equation}
\frac{D^2 \eta^\mu}{D\tau^2}= \eta^\lambda \left(\nabla_\lambda a^\mu _{(B)} g_{\nu \rho}+R^\mu _{ \lambda \nu \rho}\right)u^\rho u^\nu = \eta^\lambda \left( \mathfrak{R}^\mu_{\lambda \nu \rho}+R^\mu _{ \lambda \nu \rho}\right)u^\rho u^\nu \equiv \mathbf{R}^{\mu }_{\lambda \nu \rho}\eta^\lambda u^\rho u^\nu
\end{equation}
where, $\mathfrak{R}^\mu_{\lambda \nu \rho}=\nabla_\lambda a^\mu _{(B)} g_{\nu \rho}$ behaves like a curvature. This is only an appellation and we do not want to create sensitivity. By this definition, we do not want to say that the space-time curvature is affected in this formalism. Because, we do not consider back-reaction effects. The above relation demonstrates that when we consider non-geodesic curves or use an accelerated frame, a non-geometrical term appears alongside the space-time curvature. In this framework, we call it non-geometrical curvature.\par  In fact, the equation (\ref{rela}) can be written as $\tilde{\nabla}_\mu \tilde{\nabla}^\mu S= m_0^2$, where by $\tilde{\nabla}$ we mean that the derivatives are taken with respect to the modified metric $\tilde{g}_{\mu \nu}=\mathfrak{\omega}^2(x)g_{\mu\nu}=(1+\mathfrak{Q})g_{\mu\nu}$. In this situation, the Hamilton-Jacobi equation of the particle returns to its original form.\cite{Reffatima}.
The effect of such conformal transformation on the curvature of space-time is $ \tilde{R}^\mu_{\nu\rho\lambda}= R^\mu_{\nu\rho\lambda}+ f(\mathfrak{\omega}, \nabla_\mu \nabla_\nu \mathfrak{\omega},\cdots)$ \cite{RefWald} where $f$ is a function of conformal factor and its derivatives. This means that in our approach, the main curvature of space-time does not change and only some terms will be added to it. If we consider the back-reaction effects, then the metric $R^\mu_{\nu\rho\lambda}$ will be altered.\cite{Reffatima}. In this case some singularities of a specific space-time may be removed. But, in our approach, the main features of space-time, for example its singular points for which $\mathbf{\eta}=0 $, do not change. Because, the classical part and quantum part are separate in the deviation equation (\ref{aa3}). So, we conclude that only the trajectories and elapsed time between the two conjugate points will be changed and singular points will not be removed. In other words, the curves between two conjugate points change with respect to the classical case. \par   
An interesting point is that if we consider a flat space-time for which $R^\mu _{\rho \lambda \nu}=0$ and $\nabla_\mu \rightarrow \partial_\mu$, we get:
\begin{equation}\label{fredev}
\frac{D^2 \eta^\mu}{D\tau^2}=\eta^\lambda \partial_\lambda\left(-\frac{1}{2}u^\mu u^\nu \partial_\nu \mathfrak{Q} +\frac{1}{2}\partial^\mu \mathfrak{Q}\right)=\frac{1}{2}\eta^ \lambda \partial_\lambda (h^{\mu\nu}\partial_\nu \mathfrak{Q} )=\frac{1}{2}\eta^ \lambda \partial_\lambda a^\mu _{(\perp)}
\end{equation}
This means that even in a flat spacetime, in the presence of quantum force, geodesic deviation does not vanish. But, this deviation is not necessarily convergent. It depends on mass distribution $\rho$ and its derivatives $\partial_\mu\rho , \partial_\mu \partial_\nu \rho , \cdots $  etc, because $\mathfrak{Q}$ is dependent on density and its derivatives. In fact the relation (\ref{fredev}) reveals the non-geometrical feature of the quantum force. This is not merely a deviation for Bohmian trajectories. Rather, it is a Macian-type property which introduces the effects of mass distribution on the particle dynamics.\par 
We may face a situation in which gravitational effects is weak and velocities are very small with respect to the light velocity but quantum effects are significant. 
The relation (\ref{aa3}) in terms of quantum potential is 
\begin{equation}
\frac{D^2 \eta^\mu}{D\tau^2}=\eta^\lambda \nabla_\lambda\left(-\frac{1}{2}u^\mu u^\nu \nabla_\nu \mathfrak{Q} +\frac{1}{2}g^{\mu \nu}\nabla_\nu \mathfrak{Q}\right)+R^\mu _{\rho \lambda \nu}u^\rho  \eta ^\lambda u^\nu 
\end{equation}
At the non-relativistic limit, we can take $u^i \simeq \delta^\mu_0$, $\tau \rightarrow t$ and $R^\mu_{0\rho 0}=\partial^\mu \partial_\rho \phi(x)$, where $\phi(x)$ is the Newtonian gravitational potential. Also the time derivatives of quantum potential is taken to be zero. For weak gravitational fields, we have $\nabla_\mu \rightarrow \partial_\mu$. Then, the equation (\ref{fredev}) takes the form:
\begin{equation}\label{nredev}
\frac{\partial^2 \eta^i}{\partial t^2}=\eta^j \partial_j \left(\frac{\partial^i Q}{m_0} - \partial^i \phi(\mathbf{x}) \right)
\end{equation} 
where we have used the fact that $ln(1+Q)\simeq Q$. The relation (\ref{nredev}) can then be written more compact as:
\begin{equation}\label{qq}
\frac{\partial^2 \eta^i}{\partial t^2}=\Omega^i_j \eta^j
\end{equation}
where $\Omega^i_j$ is defined as
\begin{equation}\label{qq2}
\Omega^i_j=\partial_j \left(\frac{\partial^i Q}{m_0} - \partial^i \phi(\mathbf{x}) \right)
\end{equation}
Equations (\ref{qq}) and (\ref{qq2}) are like the equations of a oscillating system. It is obvious from the above equation that in the absence of gravitation there is always a pure quantum mechanical deviation. It may be interesting to solve some specific problems by using the above equations  for different metrics.\par 
For obtaining above relations, we started with the general relation $p_\mu p^\mu = \mathcal{M}^2$, which is true for both spin zero or spinning particles. \footnote{The Bohmian quantum potential for relativistic electrons has been derived recently by Hiley and coworkers through complicated methods of Clifford algebra $C^{0}_{3}$.\cite{RefHC}. There, the quantum potential is defined as additional terms of the Hamilton-Jacobi equation of the system and is obtained by comparing with the classical equation $p_\mu p^\mu =m_0^2$. In other words, in that case too the relation $p_\mu p^\mu =\mathcal{M}^2$ holds, where, $\mathcal{M}^2=m_0^2+Q_{Dirac}$.} Unfortunately, the quantum potential of multi-particle relativistic electron has not been obtained yet. So, the above procedure is not applicable for such systems yet. But it is possible for single-particle relativistic electron and many-particles non-relativistic electrons.
Due to the extent of the topics of spinning particle and the Clifford algebra, we can not deal with the topic of spinning electron in this paper. But here, we point out to the non-relativistic single-particle case rapidly. For a single-particle non-relativistic electron, the Bohmian quantum potential has two parts:
\begin{equation}
Q=-\frac{\hbar^2 \nabla^2 R(\mathbf{x},t)}{2m_0R(\mathbf{x},t)} +\frac{1}{2m_0}\partial_is_j \partial^i s^j , \quad i=1,2,3
\end{equation}
where $s_i$ is the ith component of the local spin vector. The local spin vector is defined in the framework of geometric algebra or Clifford algebra. We need only make a transformation $Q \rightarrow Q+Q_s$ in the relation (\ref{nredev}) to get the non-relativistic Bohmian deviation equation as:
\begin{equation}
\frac{\partial^2 \eta^i}{\partial t^2}=\eta^j \partial_j \left(\frac{1}{m_0}(\partial^i Q+\partial^i Q_s) - \partial^i \phi(\mathbf{x}) \right)
\end{equation}
The non-relativistic case is not very important. It is better to write it for single-particle relativistic case. But it is to be done in another paper.

\subsection{Deriving Bohmian Raychaudhuri's equation}
Classical Raychaudhuri's equation for a congurence of non-geodesics curves is
\begin{equation}\label{cr}
\frac{d\theta}{d\tau}=\nabla_\mu a^\mu +\omega_{\mu \nu} \omega^{\mu \nu}-\sigma_{\mu \nu} \sigma^{\mu \nu}-\frac{1}{3}\theta^2 -R_{\mu \nu}u^\mu u^\nu
\end{equation}
See ref \cite{RefPad}. So in our notation we have:
\begin{equation}
\frac{d\theta}{d\tau}=\nabla_\mu a_{(B)}^\mu +\omega_{\mu \nu} \omega^{\mu \nu}-\sigma_{\mu \nu} \sigma^{\mu \nu}-\frac{1}{3}\theta^2 -R_{\mu \nu}u^\mu u^\nu
\end{equation} 
The difference with classical Raychaudhuri's equation for a congurence of geodesics is that in latter the divergence of Bohmian acceleration is present. In terms of quantum potential we have:
\begin{eqnarray}
\frac{d\theta}{d\tau}=\frac{1}{2}\nabla_\mu \left[(g^{\mu\nu}-u^\mu u^\nu)\nabla_\nu \mathfrak{Q}\right]+\omega_{\mu \nu} \omega^{\mu \nu}-\sigma_{\mu \nu} \sigma^{\mu \nu}- \frac{1}{3}\theta^2 -R_{\mu \nu}u^\mu u^\nu = \nonumber \\ \frac{1}{2}\nabla_\mu \left[h^{\mu\nu}\nabla_\nu \mathfrak{Q}\right]+\omega_{\mu \nu} \omega^{\mu \nu}-\sigma_{\mu \nu} \sigma^{\mu \nu}- \frac{1}{3}\theta^2 -R_{\mu \nu}u^\mu u^\nu 
\end{eqnarray}
Finally, we get:
\begin{equation}\label{expq}
\frac{d\theta}{d\tau}=\frac{1}{2} \nabla_\mu \nabla^\mu \mathfrak{Q}
+\omega_{\mu \nu} \omega^{\mu \nu}-\sigma_{\mu \nu} \sigma^{\mu \nu}- \frac{1}{3}\theta^2 -R_{\mu \nu}u^\mu u^\nu
\end{equation}
Here, we have an additional term with respect to the classical Raychaudhuri's equation due to Bohmian quantum potential. For the classical case, where $\omega_{\mu\nu}=0$ and the strong energy condition is held, we conclude that the expansion must decrease during the congurence evolution. The term $\frac{1}{2} \nabla_\mu \nabla^\mu \mathfrak{Q}$ may have different signs during the evolution. This makes it difficult to judge about the behavior of trajectories during the evolution. But, from the our previous discussion about the conjugate points and deviation equation we found that since the classical curvature is not affected by Bohmian terms, the main features of space-time do not change and singularities will not be removed. On the other hand, it has been demonstrated in ref \cite{RefWald} p.225-226, that the singular points and the expansion have close and direct relationship. In other words, when $\mathbf{\eta} \rightarrow 0$, for a singular point, the geodesics expansion tends to infinity($\theta \rightarrow - \infty$). Since, in our approach the singular points will not be removed, we expect that the end of trajectories evolution leads to singular points finally. The only difference with respect to the classical procedure is that the trajectories fluctuate around the geodesics curves due to Bohmian quantum force. If, we consider the back-reaction effects by writing the full action, then some features of new space-time, like the singular points, may change.\cite{Reffatima}. Nevertheless, the consequences of singularity theorems remain unaffected. Because, singularity theorems are independent of the specific space-time whether it is classical space-time or quantum space-time.
 \par 
In the non-relativistic domain, we get:
\begin{equation}
\frac{d\theta}{dt}= -\frac{1}{m_0}\nabla^2 Q+\omega_{\mu \nu} \omega^{\mu \nu}-\sigma_{\mu \nu} \sigma^{\mu \nu}- \frac{1}{3}\theta^2 -\nabla^2 \phi(\mathbf{x})
\end{equation}
Where, $\nabla^2 \phi(\mathbf{x})=4\pi G \rho$ and $Q =-\frac{\hbar^2}{2m_0}\frac{\nabla^2 \sqrt{\rho}}{\sqrt{\rho}}$. Thus, we get:
\begin{equation}
\frac{d\theta}{dt}=\frac{\hbar^2}{2m^2_0}\nabla^2 \left(\frac{\nabla^2 \sqrt{\rho}}{\sqrt{\rho}}\right)-4\pi G \rho +\omega_{\mu \nu} \omega^{\mu \nu}-\sigma_{\mu \nu} \sigma^{\mu \nu}- \frac{1}{3}\theta^2
\end{equation}
The matter density is under the control of pilot-wave as $\rho =|\psi|^2$. It may be possible to solve these equations through the numerical methods to have a better view on quantum geodesic deviation and Raychaudhuri's equation. Specially, for the metrics of expanding universe models. 

\section{conclusion}
\label{sec:3}
We studied Bohmian deviation equation and Raychaudhuri's equation. After investigating the problem, we found that the totality of the singularity theorems and their results do not change. In other words, by using this method, we can not remove the singular or conjugate points of space-time. Because, the output of this method is the sum of old space-time features with additional features due to the Bohmian quantum potential that are not correlated to each other. Thus, the main features of classical curvature e.g. singular points, horizons and etc, remain unaffected and what changes is a deviation from geodesic equation. Only in one case the singularities of space-time may be removed, and it is when we consider the effects of back-reaction of matter on space-time in the Bohmian framework. These results are obtained in the framework of BQM in which the definition of trajectory is possible. Otherwise, it was not possible to get these results through the SQM.




\end{document}